\documentclass[10pt,A4]{article}
\usepackage[top=0.85in,left=0.85in,footskip=0.75in,marginparwidth=2in]{geometry}

\usepackage{moreverb}
\usepackage[colorlinks,bookmarksopen,bookmarksnumbered,citecolor=red,urlcolor=red]{hyperref}
\usepackage[textwidth=8em,textsize=small]{todonotes}
\usepackage{amsmath}
\usepackage{changes}
\usepackage{rotating}
\usepackage{bigstrut}
\usepackage{graphicx}

\usepackage[utf8]{inputenc}

\usepackage{cite}

\usepackage{nameref,hyperref}

\usepackage{changepage}

\usepackage[aboveskip=1pt,labelfont=bf,labelsep=period,singlelinecheck=off]{caption}

\makeatletter
\renewcommand{\@biblabel}[1]{\quad#1.}
\makeatother

\usepackage{lastpage,fancyhdr,graphicx}
\usepackage{epstopdf}
\fancyheadoffset[L]{2.25in}
\fancyfootoffset[L]{2.25in}

\usepackage{color}

\definecolor{Gray}{gray}{.25}

\usepackage{wrapfig}
\usepackage[pscoord]{eso-pic}
\usepackage[fulladjust]{marginnote}
\reversemarginpar

\begin{document}
\vspace*{0.35in}

\begin{flushleft}
{\Large
\textbf\newline{Improving phase II oncology trials using best observed RECIST response as an endpoint by modelling continuous tumour measurements}
}
\newline
\\
Chien-Ju Lin\textsuperscript{1,*},
James M.S. Wason\textsuperscript{1},
\\
\bigskip
\bf{1} MRC Biostatistics Unit, Cambridge, UK
\\
\bigskip
* chienju@mrc-bsu.cam.ac.uk

\end{flushleft}

\section*{Abstract}
In many phase II trials in solid tumours, patients are assessed using endpoints based on the Response Evaluation Criteria in Solid Tumours (RECIST) scale. Often, analyses are based on the response rate. This is the proportion of patients who have an observed tumour shrinkage above a pre-defined level and no new tumour lesions. The augmented binary method has been proposed to improve the precision of the estimator of the response rate. The method involves modelling the tumour shrinkage to avoid dichotomising it. However, in many trials the best observed response is used as the primary outcome. In such trials, patients are followed until progression, and their best observed RECIST outcome is used as the primary endpoint. In this paper, we propose a method that extends the augmented binary method so that it can be used when the outcome is best observed response. We show through simulated data and data from a real phase II cancer trial that this method improves power in both single-arm and randomised trials. The average gain in power compared to the traditional analysis is equivalent to approximately a 35$\%$ increase in sample size. A modified version of the method is proposed to reduce the computational effort required. We show this modified method maintains much of the efficiency advantages.



\section{Introduction}
\label{sec1}

A new cancer treatment is tested for potential benefit in phase II trials that use a relatively small number of patients followed over a short period of time. The results of the phase II trial determines whether to test the treatment in a larger, more time-consuming, and more costly phase III trial. Because of the high cost of, and high failure rate in, phase III oncology trials \cite{Paul2010}, it is important to improve the analysis of Phase II trials to ensure the decision is more accurate.

Phase II oncology trials use a variety of endpoints to evaluate the efficacy of a treatment \cite{Johnson2003,Pazdur2008}. The most commonly used endpoints are based on the Response Evaluation Criteria in Solid Tumours (RECIST) scale \cite{Eisenhauer2009}. RECIST defines tumour size as the sum of longest diameters of target lesions and categorises patients into complete response (CR), partial response (PR), stable disease (SD), and progressive disease (PD). CR and PR represent no new tumour lesions and a 100$\%$ shrinkage and greater than 30$\%$ shrinkage, respectively; PD represents a 20\% increase in tumour size from the minimum size observed up to that point, or new lesions appearing. Often patients are followed until they are categorised as PD or a preplanned time, and patients with CR or PR are labelled responders. The response rate is defined as either: 1) proportion of patients who are responders at a certain time after baseline (fixed response); or 2) the proportion of patients whose best observed response before progression is CR or PR (best observed response, BOR).

Categorizing patients into responders and non-responders is widespread and clinically appealing. However, it can have substantial statistical disadvantages. Its major limitation is that it dichotomises the continuous tumour variable, thus discarding information. This loses substantial efficiency \cite{Dhani2009}. Some researchers have addressed the problem and proposed methods to make use of the continuous change in tumour response to improve statistical efficiency. This is done in different ways. Karrison \textit{et al.} \cite{Karrison2007} propose directly using the change in tumour size as an endpoint. Wason and Seaman \cite{Wason2013} use models for the tumour size and new lesion data which can be used to infer the fixed response rate with higher precision. Jaki \textit{et al.} \cite{Jaki2013} propose a method that links tumour size change with mortality using historical datasets. Authors have demonstrated that using continuous scales can increase the power (or reduce the required sample for a target power) compared to analysing the binary composite outcome.

The method of Wason and Seaman \cite{Wason2013} retains the clinically meaningful endpoint but takes into account the continuous information on tumour size. This method is limited by only allowing two follow-up visits, and only considering response rate at not only a fixed time (i.e. it cannot be used to make inferences on BOR). In trials that patients are assessed two times (interim and final), their method is sufficient. However, in trials that patients are followed up until a pre-planned time, a method incorporates information on all measurement data is preferred. In this paper we present an extended method that can be used for any number of follow-up times for fixed response or BOR. We propose a modified version that uses a highly efficient technique for multivariate integration \cite{Genz2009}, which substantially reduces the computation time taken. We assess the properties of the proposed methods by using simulated data and data from a real phase II cancer trial (HORIZON II).

This paper is divided into four sections. Section \ref{sec2} gives a brief overview of the augmented binary method \cite{Wason2013}. It then describes the proposed extensions of the method. Section \ref{simulation} evaluates the performance of the proposed methods using simulations and real data. Section \ref{sec4} summarizes the results and presents limitations and future work.


\section{Methods}
\label{sec2}
\subsection{Background}

We use the phrase 'tumour size' as shorthand for the sum of the longest diameter of target tumour lesions. We assume patients’ tumour sizes are recorded until progression occurs or until a preplanned number of visits. We note there are two ways in which a progression can occur: an increase in tumour size by more than 20 \% (a tumour-growth progression) or new lesions appearing (a new-lesion progression). Two response endpoints can be used in the analysis, one being \textit{fixed time} and the other being BOR. Analysis at a \textit{fixed time} $t$ uses the proportion of responders at time $t$ (those who have a tumour size shrinkage at time $t$ above a pre-defined threshold and no progression up to that point). BOR defines patients as a responder or not according to their best observed response before progression. The latest RECIST guidelines \cite{Eisenhauer2009} give BOR two definitions according to whether confirmation is required or not. Confirmation means that an apparent response must be backed up by continued response at the next timepoint to be counted as genuine. This is especially recommended for single-arm trials. When confirmation is not required (randomised trials comparing two arms), BOR is defined as the best response across all time points up to progression. When confirmation is required, BOR is defined as a response if the patient is a responder at two consecutive time points before progression.

\subsection{Notation}

Tumour sizes for each patient are measured at several discrete times ($T$ denoting the maximum time). The tumour size at time $t$ for patient $i$ is denoted by $z_{it}$ where $t=0$ represents the baseline measurement. We denote $G$ and $X$ as the time at which a tumour-growth progression and new-lesion progression occurs, respectively. Once a patient progresses they are no longer followed up. The observed data is therefore $(F_i, \textbf{z}_i)$ where $F_i$=min$(X_i,G_i,T)$. We define $y_{it}$ as the log tumour size ratio for patient $i$ at time $t$, $y_{it} =\log(z_{it}/z_{i0})$, and $c_t$ as the pre-specified dichotomisation threshold for response (on the log tumour ratio scale). Further, $D_{it}$ defines new-lesion progression indicators: \{$D_{it}=1$ if patient $i$ has a progression due to new lesions occurring between time $(t-1)$ and $t$, $t=1,\ldots,T$ \}. We define composite response indicators corresponding to the definitions of \textit{fixed time} and BOR as follows.

For \textit{fixed time} at time $t$, the composite response indicator $S_{it}$ for patient $i$ is defined as
\[
\begin{array}{r}
S_{it}= \left\{\begin{array}{ll}
                1, & \text{ if $D_{ij}=0$ for all $j=1,\ldots,t$ and $y_{it} < c_t$} \\
                0, & \text{ otherwise}\\
                \end{array} \right.
\end{array}.
\]

For BOR when confirmation is not required, the event is equivalent to having at least one record classified CR/PR before progression or time $T$, the response indicator $BOR_i$ is defined as
\[
\begin{array}{r}
BOR_i= \left\{\begin{array}{ll}
                1, & \text{ if there exists a $t$ such that } y_{it}<\log(0.7),t\leq min(\text{F}_i,X_i-1)\\
                0, & \text{ otherwise}\\
                \end{array} \right.
\end{array}.
\]

We consider the case where confirmation is required later.

\subsection{Estimating response probability using the augmented binary method with two follow-up times} \label{Augbin}
The augmented binary method, henceforth referred to as AugBin, was proposed by Wason and Seaman \cite{Wason2013}. We briefly describe this method here, but more details are found in \cite{Wason2013}.

The AugBin method makes assumptions that the log tumour size ratios follow a multivariate normal distribution, and the probability of new-lesion progression depends only on the observed tumour size at the previous visit. The log tumour size ratios are modelled by
\[
(\text{Y}_{i1}, \text{Y}_{i2})'|z_{i0} \sim \hbox{N}((\mu_{i1}, \mu_{i2})',\Sigma),
\]
and the new-lesion progression is modelled by using logistic regression models
\[
\hbox{Logit}\{\text{Pr}(D_{i1}=1|z_{i0})\}=\alpha_1+\gamma_1z_{i0},
\]
\[
\hbox{Logit}\{\text{Pr}(D_{i2}=1|D_{i1}=0,z_{i0},z_{i1})\}=\alpha_2+\gamma_2z_{i1}.
\]
The probability of response for patient $i$ is written by:
\[
\text{Pr}(S_i=1|\theta)=\int_{-\infty}^{c_2}\int_{-\infty}^{\infty}\text{Pr}(D_{i1}=0|z_{i0})\text{Pr}(D_{i2}=0|D_1=0,z_{i0},z_{i1})f_{Y_1Y_2}(y_{i1}y_{i2};\theta)dy_{i1}dy_{i2},
\]
where $\theta$ is the vector of parameters from the above models and $c_2$ is the dichotomisation threshold (usually log(0.7), representing at least a 30$\%$ shrinkage in the tumour size from baseline). The mean response probability is estimated by $\overline{\text{Pr}}(S=1|\hat{\theta})=\frac{1}{n}\sum_{i=1}^{n}\text{Pr}(S_i=1|\hat{\theta})$, where $\hat{\theta}$ is the maximum likelihood estimator of $\theta$. A program  is available in the paper which uses R2Cuba to compute the above integration. An approximately (1-$\alpha$) \%  confidence interval for the probability of response is constructed on the logit scale, that is, expit$\bigg\{l(\hat{\theta})\pm \Phi^{-1}(1-\frac{\alpha}{2})\sqrt{\text{var}(l(\hat{\theta}))}\bigg\}$, where $l(\theta)=\frac{\text{Pr}(S_i=1|\theta)}{1-\text{Pr}(S_i=1|\theta)}$ and $\text{var}(l(\hat{\theta}))$ is obtained by using the delta method.

\subsection{Extended augmented binary method at a fixed time (t$>$2)} \label{modifyAug}
We use the same assumptions and extend the AugBin method to $t$ follow-up times. The log tumour size ratios are modelled by
\begin{equation}
(\text{Y}_{i1}, \dots, \text{Y}_{it})'|z_0 \sim \hbox{N}((\mu_{i1}, \ldots, \mu_{it})',\Sigma).
\end{equation}
An unstructured covariance matrix is used (although an alternative form may be needed if $t$ is large enough). The new-lesion progression is modelled by
\begin{equation}
\hbox{Logit}\{\text{Pr}(D_{it}=1|D_{i1}=\ldots=D_{i(t-1)}=0 , z_{i0},\ldots,z_{i(t-1)})\}=\alpha_t+\gamma_tz_{i(t-1)}.
\end{equation}
We assume that the new-lesion progression depends only on the previous observed rumour size. The missing tumour size because of new-lesion progression can be, therefore, treated as MAR as justified in \cite{Wason2013}. We assume data $(Y_{i1},\ldots,Y_{iF_i})$ before progression is always observed and $(Y_{iF_{i+1}},\dots,Y_{iT})$ is always missing, similarly for $(D_{i1},\ldots,D_{iF_i})$ and $(D_{iF_{i+1}},\dots,D_{iT})$. The data $(Y_1,\dots,Y_T, D_1,\dots,D_T)$ is the case of monotone missingness. The probability of response for patient $i$ at time $T$ can be written by:
\begin{equation}
\begin{array}{cl}
\text{Pr}(S_{iT}=1|\theta)=\int_{-\infty}^{c_T}\ldots \int_{-\infty}^{\infty} &\prod\limits_{t=1}^{T}\text{Pr}(D_{it}=0|D_{i1}=\ldots=D_{i(t-1)}=0 , z_{i0},\ldots,z_{i(t-1)})\\
 & \times f_{Y_1,...,Y_T}(y_{i1},...,y_{iT};\theta)dy_{i1}...dy_{iT}
  \end{array}
  \label{eq.eAugbin}
\end{equation}

The advantage is that the Equation (\ref{eq.eAugbin}) uses the models to estimate probability of response of patients and missing data are MAR, it can be applied to patients who drop out before preplanned time. The probability is interpreted as the probability of patient $i$ being a responder at time T as if they were observed until T. A potential issue of Equation (\ref{eq.eAugbin}) is that the multivariate integration is computationally intensive. The mean response probability is estimated by averaging response probability over $n$ patients given $\hat{\theta}$. An approximately (1-$\alpha$) \% confidence interval is constructed as described in \ref{Augbin}.

\subsection{Modified augmented binary method at a fixed time}\label{mAugbinfix}
The objective for this section is to efficiently estimate the mean response probability using continuous tumour-size information in a computationally efficient way. We assume that \{no new-lesion progression occurs from time 1 to time T\} and  \{no tumour-growth progression\} are conditionally independent given tumour size $\bar{z}_{t-1}=(z_0,\ldots,z_{t-1})$. We note this is a strong assumption, and assess the sensitivity to this assumption later on. The probability of response for patient $i$ at a \textit{fixed time} $t$ can be written by
\[
\begin{array}{cl}
\text{Pr(response$|\bar{z}_{t-1}$)} & = \text{Pr(no new-lesion progression until $t| \bar{z}_{t-1}$)} \\
& \times \text{Pr(no tumour progression at time $t | \bar{z}_{t-1}$ )}.
\end{array}
\]
Let $\pi_{t}$ be the probability of new-lesion progression at time $t, t=1, \dots, T$. Note that $\pi_t$ is a conditional probability given no new-lesion progression occurring at previous timepoints. The log tumour size ratio $\mathbf{Y}_i$ is allowed to depend on baseline tumour size whereas new-lesion progression depends on the previous observed tumour size at the previous visit. We can model Y by
\begin{equation}
(\text{Y}_{i1}, \dots, \text{Y}_{it})'|z_0 \sim f_{Y_i}(\cdot), l(\pi_{it})=\beta_0+\beta_t z_{i(t-1)}.
\end{equation}
where $f_Y(\cdot)$ is a joint distribution, and $l(\cdot)$ is the logit link function. We assume that $\pi_{it}(\bar{z}_{t-1})=\pi_{it}(z_{t-1})$. The probability of response for patient $i$ at a \textit{fixed time} $T$ can be written by

\begin{equation}
\text{Pr}(S_{i}=1|\bar{z}_{t-1}, \theta)=\prod_{t=1}^{T} \{1-\pi_{it}(\bar{z}_{t-1},\theta)\} \int_{-\infty}^{c_T}\int_{-\infty}^{\infty} ... \int_{-\infty}^{\infty} f_{Y_1,...,Y_T}(y_{i1},...,y_{iT}|z_{i0}, \theta)dy_{i1}...dy_{iT},
\label{eqfix}
\end{equation}
where $c_T$ is the dichotomisation threshold and $\theta$ is a vector of parameters of the models. We assume that $f_Y(\cdot)$ is the pdf of a multivariate normal distribution. The multivariate integration can then be calculated by a highly efficient technique proposed by Genz and Bretz \cite{Genz2009}. The value of $\pi_{it}$ is estimated by using $l(\hat{\pi}_{it})$ if $z_{i (t-1)}$ is observed. For patients who have progressed, their records at time $(F_i+1),\ldots, T$ are not observed. Their probability of new-lesion progression at time $t, t> F_i$ is estimated by
\begin{equation}
\tilde{\pi}_{it}=\frac{1}{n_t-k}\sum \limits_{j,y_{jt} \in \varphi}{}\pi_{jt}
\label{eqpi2}
\end{equation}

where $n_t$ is the number of patients with observed $z_{t-1}$ and $k$ is the number of patients who have log tumour size ratio $y_t$ outside of the region of integration $\varphi$ of Equation (\ref{eqfix}). We trim those $k$ patients to avoid underestimating $\pi_t$. This is similar to an idea of trimmed mean, which is used in many areas and has advantages under both normal and non-normal distributions \cite{Stigler1973, Wilcox2005}.

The vector $\theta$ consists of \{(T+1)+ T(T+1)/2 + 2T \} parameters (the $\mu$ and $\Sigma$ parameters from the multivariate normal and parameters from the logistic regression models). The mean response probability is estimated by  $\overline{\text{Pr}}(S=1|\hat{\theta})=\frac{1}{n}\sum_{i=1}^{n}\text{Pr}(S_i=1|\hat{\theta})$, where $\hat{\theta}$ is the maximum likelihood estimator of $\theta$. A $(1-\alpha)$ \% confidence interval for $\overline{\text{Pr}}(S=1|\theta)$ can be constructed :
\[
\Bigg[\overline{\text{Pr}}(S=1|\hat{\theta})-\Phi^{-1}(1-\alpha/2)\sqrt{\text{Var}(\overline{\text{Pr}}(S=1|\hat{\theta}))} , \; \overline{\text{Pr}}(S=1|\hat{\theta})+\Phi^{-1}(1-\alpha/2)\sqrt{\text{Var}(\overline{\text{Pr}}(S=1|\hat{\theta}))}\Bigg],
\]
where $\Phi$ is the standard normal distribution function. However, we found that the method has better properties if we find a confidence interval for logit$\{\overline{\text{Pr}}(S=1|\hat{\theta})\}$ and transform back. Let $l(\theta)=\log \frac{\text{Pr}(S_i=1|\theta)}{1-\text{Pr}(S_i=1|\theta)}$, we obtain $\text{Var}(l(\hat{\theta}))$ by using the delta method, which is written by
\[
\text{var}(l(\hat{\theta})) \approx (\nabla l(\hat{\theta}))^T \text{var}(\hat{\theta}) \nabla l(\hat{\theta}),
\]
where $\nabla l(\hat{\theta})$ is the partial derivatives of $l(\theta)$. An approximately (1-$\alpha$) \%  confidence interval for the probability of response is
\[
\Bigg[\text{expit}\bigg\{l(\hat{\theta}) - \Phi^{-1}(1-\alpha/2)\sqrt{\text{var}(l(\hat{\theta}))}\bigg\}, \; \text{expit}\bigg\{l(\hat{\theta})+ \Phi^{-1}(1-\alpha/2)\sqrt{\text{var}(l(\hat{\theta}))}\bigg\}\Bigg].
\]

To summarise, the modified method uses a simplification for the relationship between new-lesion progressions and tumour-growth progressions in order to use a more efficient procedure for multivariate integration.

\subsection{Proposed method for best observed response (BOR)}\label{BOR}
We focus on the case where confirmation is not required but show briefly how the methodology can straightforwardly allow for confirmation later. By the definition of BOR, a patient is a responder if they have at least one log tumour size ratio smaller than log(0.7) before progression or maximum follow-up time. We define $\Omega_1= (\log(0.7), \log(1.2))$, $\Omega_2 = (-\infty,\log(0.7))$, and $\Omega_3=(-\infty, \infty)$ as the possible regions of integration corresponding to being classified as stable disease, responder, and irrelevant variables. Let $h$ be the time at which the patient is first classified as CR/PR. Hence, each component of $\bar{Y}_T$ will fall into one of the three regions as
\begin{equation}
(Y_1 \ldots Y_{h-1} \in \Omega_1, Y_h \in \Omega_2, Y_{h+1} \ldots Y_T \in \Omega_3).
\label{BORy}
\end{equation}
The probability of response using BOR for patient $i$ will be the sum over all possibilities of when the CR/PR is first observed. Following the concept of the extended augmented binary approach (eAugbin), the probability of response can be written by :
\begin{equation}
\begin{array}{cl}
\text{Pr}(BOR_i=1|\theta)= \sum \limits_{h=1}^T \int\limits_{\Omega_3^{T-h}}\int\limits_{\Omega_2^1} \int\limits_{\Omega_1^{h-1}} &\prod\limits_{t=1}^{h}\text{Pr}(D_{it}=0|D_{i1}=\ldots=D_{i(t-1)}=0 , z_{i0},\ldots,z_{i(t-1)})\\
 & \times f_{Y_1,...,Y_T}(y_{i1},...,y_{iT};\theta)dy_{i1}...dy_{iT}.
  \end{array}
\end{equation}

Similarly, following the concept of the modified augmented binary approach (mAug), the probability of response can be written by:
\begin{equation}
\text{Pr}(BOR_i=1|\bar{z}_{t-1}, \theta)= \sum \limits_{h=1}^T\prod_{t=1}^h \{1-\pi_{it}(\bar{z}_{t-1}, \theta)\} \int\limits_{\Omega_3^{T-h}}\int\limits_{\Omega_2^1} \int\limits_{\Omega_1^{h-1}} f_{Y_1,...,Y_T}(y_{i1},...,y_{iT};z_{i0}, \theta)dy_{i1}...dy_{iT}.
\label{eqBOR}
\end{equation}

The mean response probability is then estimated by $\frac{1}{n}\sum_{i=1}^{n}\text{Pr}(BOR_i=1|\hat{\theta})$, where $\hat{\theta}$ is the maximum likelihood estimator of $\theta$. As before, we work on the logit scale, use the delta method to obtain the variance, and then transform back to construct the confidence interval for the mean response probability.

When confirmation is required, having two continued responses of CR/PR before progression, one can replace (\ref{BORy}) with $(Y_1 \ldots Y_{h-1} \in \Omega_1, Y_h, Y_{h+1} \in \Omega_2, Y_{h+2} \ldots Y_T \in \Omega_3)$ with the sum in (\ref{eqBOR}) going from 1 to $T-1$.


\subsection{Testing a difference in probability of response between two treatments}\label{twoarms}
The above methods can be applied to single-arm trials. For a randomised trial where comparing the difference in response probability is of interest, a minor addition is required.

We assume $2n$ patients are recruited with n patients randomised to each arm. Assumptions for log tumour size ratios and new-lesion progression remain the same as in Section \ref{modifyAug}. We introduce an arm indicator R to the models, with 0 for control and 1 for experimental arms. The log tumour size ratios are modelled by
\[
(\text{Y}_{i1}, \text{Y}_{i2},\ldots, \text{Y}_{iT})'|\text{R},z_0 \sim \hbox{N}((\mu_{i1}, \mu_{i2}, \ldots, \mu_{iT})',\Sigma),
\]
the new-lesion progression for $T>t$ is modelled by using logistic models
\[
\hbox{Logit}\{\text{Pr}(D_{it}=1|D_{i1}=\ldots=D_{i{t-1}}=0, z_{i0},\ldots,z_{i(t-1)})\}=\alpha_t+\beta_t\text{R}+\gamma_tz_{i(t-1)}.
\]
The probabilities of new-lesion progression for control and experimental arms are $[1+\exp \{-(\alpha_t+\gamma_tz_{i(t-1)})\}]^{-1}$ and $[1+\exp \{-(\alpha_t+\beta_t+\gamma_tz_{i(t-1)})\}]^{-1}$, respectively. Let $\theta$ be the vector of $(T^2+ \frac{9T }{2}+2)$ parameters from the above models.
The mean response probability at a \textit{fixed time} is estimated by
\[
\overline{\text{Pr}}(S=1|\hat{\theta},\text{R})=\frac{1}{2n}\sum\limits_{i=1}^{2n}\text{Pr}(S_i=1|\hat{\theta},\text{R}),
\]
where $\hat{\theta}$ is the maximum likelihood estimator of $\theta$. We note that patients from both arms are included in the calculation of the probability of response in an arm, as is recommended and justified in \cite{Wason2013}. The mean difference in response probability at a \textit{fixed time} is defined as the difference between mean response probabilities for the two arms. It can be written by
\[
\text{m}_F(\theta)=\overline{\text{Pr}}(S=1|\text{R}=1,\theta)-\overline{\text{Pr}}(S=1|\text{R}=0,\theta).
\]
We obtain the variance of $\text{m}_F(\hat{\theta})$ by using the delta method and use the Wald test to test whether $\text{m}_F(\theta)$ is zero. Similarly, we define the mean difference in response probability for BOR as
\[
\text{m}_\text{B}(\theta)=\overline{\text{Pr}}(BOR_i=1|\text{R}=1,\theta)-\overline{\text{Pr}}(BOR_i=1|\text{R}=0,\theta).
\]

Both the extended and modified methods can be used as in previous sections.

\section{Results}\label{simulation}
In this section, we evaluate the performance of the proposed methods in terms of precision and power using simulations and a real data example. We use ``Bin'' to represent the method that just analyses the response outcomes as binary. For single-arm trials, the binary method uses the R-package Hmisc to construct a Wilson interval for binary success $(S=1$ or $BOR=1)$. For two-arm studies, the binary method is a logistic regression model that has parameters for treatment group and baseline tumour size, from which the treatment effect can be tested. The terms ``Augbin'', ``eAugbin'', and ``mAug'' refer to methods that use continuous information. They are, respectively, Wason and Seaman's method \cite{Wason2013} at two-follow up times, the extended method for more than two-follow up times, and the modified method for rapid computation. We use \textit{fixed time} with varying numbers of follow-up times and best observed response without confirmation as the endpoints.

\subsection{Simulation study setup}
Following the aforementioned notation, the observed data available for each patient is $(F_i, \textbf{z}_i)$. The observed data are simulated as follows. First of all, baseline tumour size $z_{i0}$ for patient $i$ is generated from a uniform distribution and log tumour size ratios of T follow-up time $\{y_{it}: t=1,\dots,T \}$ are generated from a multivariate normal distribution. Tumour size $z_{it}$ can then be calculated from $z_{it}=e^{y_{it}}z_{i0}$. Next, new-lesion progression indicators are generated from logistic models with intercept $\alpha$ and tumour size effect $\gamma$. A non-zero $\gamma$ means that probability of new-lesion progression depends on the tumour size at the previous timepoint. We define time to new-lesion progression as the first time when the new-lesion progression occurs from the logistic models. Finally, tumour size observations of patient $i$ after progression are replaced as missing. 

\subsubsection{Single-arm trials assessing response at fixed time}\label{simulation-singlearm} ~\\ 
Before generating 5000 replicates, we test the computation time for running one replicate using Augbin/eAugbin. We generated one replicate of 75 patients. Baseline tumour size $(Z_0)$ is generated from a uniform distribution and log tumour size ratios are generated from a multivariate normal distribution for 2 to 6 follow-up times. The $(\alpha, \gamma)$ are set to (-1.5,0) and (-2.5,0.2). The value of $\alpha=-1.5$ corresponds to an 18 \% chance of developing new lesions between each visit. The computation time for running one replicate using Augbin/eAugbin for 2 to 6 follow-up times are 0.04, 0.65, 2.28, 3.41 and 4.47 minutes; while mAug at 6 follow-up times takes 0.09 minutes. We do not consider $T>4$ because of the length of time need to simulate 5000 replicates for eAugbin. The simulation settings of log tumour size ratios for 2 follow-up times is a similar formulation to \cite{Wason2013}, that is
\begin{align*}
Z_0 \sim U(0,1),\textbf{Y}_{2}\sim  \text{N}
\begin{bmatrix}
\log(0.7)
\begin{pmatrix}
.5\\
1
\end{pmatrix}\!\!,
\begin{pmatrix}
.5 & .5 \\
.5 & 1
\end{pmatrix}
\end{bmatrix}.\\
\end{align*}

The settings for T=3 and 4 are: $Z_0 \sim U(0,1),$
\begin{align*}
\textbf{Y}_{3}\sim  \text{N}
\begin{bmatrix}
\log(0.7)
\begin{pmatrix}
.5\\
.75\\
1
\end{pmatrix}\!\!,
\begin{pmatrix}
.5 & .5 & .5\\
.5 & .75 & .75\\
.5 & .75 & 1
\end{pmatrix}
\end{bmatrix}
,\textbf{Y}_{4}\sim  \text{N}
\begin{bmatrix}
\log(0.7)
\begin{pmatrix}
.25\\
.5\\
.75\\
1
\end{pmatrix}\!\!,
\begin{pmatrix}
.25 & .25& .25& .25\\
.25 & .5 & .5 & .5\\
.25& .5 & .75 & .75\\
.25& .5 & .75 & 1
\end{pmatrix}
\end{bmatrix}.\\
\end{align*}

Table \ref{Table2} shows mean estimated response probability and coverage for Bin, Augbin/eAugbin and mAug for 2,3,4 follow-up times for 5000 replicates. The columns 10-11 show the reduction in 95\% confidence interval (CI). They are, respectively, the average of [1-(CI width of Augbin)/(CI width of Bin)] and [1-(CI width of mAug)/(CI width of Bin)]. As seen, in all cases, eAugbin and mAug have narrower CIs compared with Bin. For example, mAug reduces CI width by 14\% means that Bin needs an additional 30\% sample size to obtain a similar width. The mAug has a similar coverage to Augbin at $t=2$. For larger t, it appears the mAug method has a better coverage probability (i.e. closer to the nominal value) than eAugbin. The reduction in confidence interval width, compared to the binary method, appears to be similar for the two methods. Thus for single-arm trials it appears mAug shows a significant improvement in computational efficiency without notably poorer statistical characteristics compared to eAugbin.\\

\subsubsection{Randomised trials using response at fixed time} ~\\
We consider a two-arm trial with a control and experimental arm for 2 follow-up times. Each arm has 75 patients that have been allocated at random. Baseline $(Z_0)$ is generated from a $U(0,1)$ distribution. The mean log tumour size ratios between each visit are generated from a normal distribution with mean $\mu$ and variance $\frac{1}{2}$. We set $\mu= \log(0.7) + \delta \tau + \psi$, where $\delta=1$ for control and $\delta=-1$ for experimental arms, $2\tau$ is the difference in the mean log tumour size ratio and $\psi$ reflects the effectiveness of the control treatment. This is a similar formulation as \cite{Wason2013}.

\begin{figure}
\centering
\includegraphics[scale=0.7]{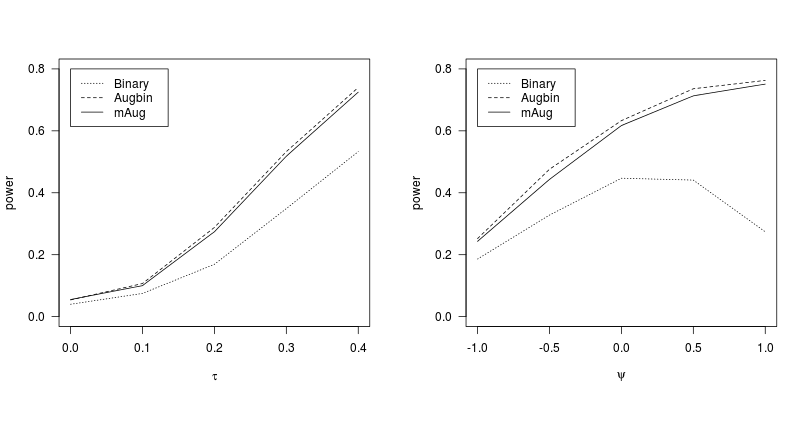}
\caption{\label{Fig1}Power of the three methods for \textit{fixed time} as the mean log tumour size ratio $(\tau)$ varies and as $\psi$ varies at $\tau =.35$.}
\end{figure}

Figure \ref{Fig1} compares the powers for Bin, eAugbin and mAug methods for randomised trials. The figure on the right shows the power over treatment effect when $\tau$ =.35. As seen, there is a clear power gain when using either mAug or Augbin. mAug performs very closely to Augbin. The empirical Type I error when the difference is 0 for Augbin and mAug are 0.054 and 0.055, respectively.

\subsubsection{Non-comparative trials for BOR} ~\\
Using the binary composite outcome, patients are classified as responders if they have a CR/PR before time F. The computation time for running one replicate using Augbin/eAugbin and BOR for 3 to 6 follow-up times are 0.05, 0.09, 0.3 and 0.56 minutes; while mAug at 6 follow-up times takes 0.22 minutes. Again we use 5000 replicates of 75 patients. Baseline tumour size is generated from a uniform distribution (0, 1). The log tumoursize ratios are generated from multivariate normal distribution for 4,5,6,7 follow-up times with $\sigma_{tt}^2=1, t=4,5,6,7$. Regardless of the number of visits after baseline, we set the mean log tumour size ratios at the end of the treatment to $\log(0.7)$. For example,  the case where T=4 refers to having 4 visits after baseline and $\pmb{\mu}$ being set to $\log(0.7)$ [$.25 \: .50 \: .75 \: 1$]. For computational reasons, eAugbin was included for up to T=6. Tables \ref{Table3} show the operating characteristics of mAug and Bin for maximum number of visits varying from 4 to 7. Overall, mAug reduces the average width of the CI by at least 16 \% compared with Bin.  This is equivalent to needing a sample size of around 101 $(1.16^2 \times 75)$, to obtain a similar average width using Bin. The reduction in width is slightly higher when there is a tumour size effect on new-lesion progression.

\subsubsection{Comparative trials for BOR} 
\begin{figure}
\centering
\includegraphics[scale=0.5]{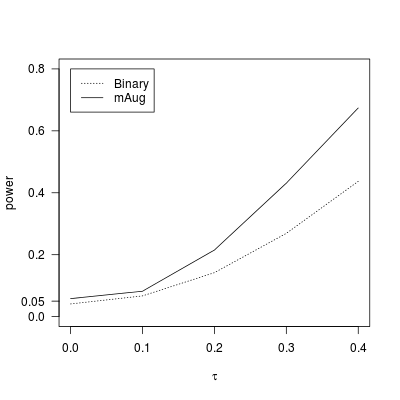}
\caption{\label{Fig2} Power of the binary method and mAug for best observed response for 4 time points as the mean log tumour size ratio $(\tau)$ varies.}
\end{figure}

~\\
To illustrate results of the mAug method for a two-arm trial, we consider the case where each arm has 75 patients and patients are followed for 4 time points. The mean log tumour size ratios for each time point is $( \log(0.7) + .25\delta \tau)$, $( \log(0.7) + .5\delta \tau)$, $( \log(0.7) + .75\delta \tau)$, and $( \log(0.7) +\delta \tau)$, where $\delta=1$ for control and $\delta=-1$ for experimental arms respectively. Figure \ref{Fig2} compares the powers for Bin and mAug methods in comparative trials for four time points when best observed response is used. Although there is a slight inflation in Type I error rate for mAug, in general, there is a consistent power advantage when using mAug compared to using Bin. The empirical Type I error when the difference is 0 for Binary and mAug are 0.041 and 0.058, respectively.

\subsection{Case study: HORIZON II}\label{Horizon}

HORIZON II (clinicaltrials.gov identifier: NCT00384176) is a three arm colon cancer trial sponsored by AstraZeneca. Patients initially were randomly assigned 1:1:1 to placebo, cediranib 20 mg once daily, cediranib 30 mg once daily. Later, subsequent patients were randomly assigned 1:2 to placebo or cediranib 20 mg \cite{Hoff2012}. The numbers of patients with baseline record for the three arms are 346, 484, 209, respectively. The tumour sizes of patients were measured every 6 weeks up to 24 weeks and then every 12 weeks. Figure S1 in supporting information shows a waterfall plot for the individual reduction in tumour size at week 24 from the baseline. There are cases that participants are classified as responders before progression which results in different response estimates between \textit{fixed time} and BOR.

We used a permutation test to calculate the empirical type I error rate. Data from baseline, 6, 12, 18, and 24 weeks were used. We simulated 5000 replicates, with the treatment assignment label shuffled randomly in each replicate, For each replicate, we tested the difference in probability of best observed response between two treatment arms using mAug with 4-follow up times. The empirical Type I error for no difference between placebo and cediranib 20 mg is 0.0558 and that between placebo and cediranib 30 mg is 0.0518. These are within Monte Carlo standard error of a true type I error of 0.05 (MC error +/- 0.006).

Figure S2 in supporting information shows the mean estimated response probability using the three methods and \textit{fixed time} with between 2 and 5-follow up times for Placebo, 20 mg and 30 mg, respectively. The mean estimated response probability decreases as the number of timepoints increases. Generally, the estimated mean probabilities of response for three methods are similar.

Table \ref{Table5} reports the width of the 95\% CI for each arm’s probability of response using \textit{fixed time}. The width corresponds to the length of the vertical lines shown in Figure S2. The 95\% CI widths of eAugbin and mAug are considerably narrower than that of Bin. We compared Placebo and cediranib 20 mg as well as Placebo and cediranib 30 mg using mAug BOR and Bin BOR for 4-6 time points. Results show that the mAug method gives a considerably smaller 95\% CI than the Bin method. The maximum width of the 95\% CI for mAug is 0.131 for comparing Placebo with 30 mg, while the width is 0.174 for Bin (See Table S2 in supporting information).

\section{Discussion}\label{sec4}
In this paper we have considered how the augmented binary method of \cite{Wason2013} can be extended to be applicable for a wider range of phase II oncology trials. We have made three contributions. The first is to extend the existing method to more than two follow-up times. The second is a modified method that considerably reduces the computational time by making a simplifying assumption about the relationship between new lesions and tumour size change. The third is a mechanism for using both of these methods when the endpoint is based around the best observed RECIST observation before progression, which is a common phase II oncology endpoint.

We have shown that all proposed methods carry the same good properties as the augmented binary method. They provide extra precision, i.e. they require a smaller sample size for the same precision (compared to the traditional analysis of analysing response as a binary outcome) in single arm trials and are more powerful in comparative trials.

The difference between the modified(mAug) and extended method (eAugbin) is that the former uses the estimated probability of new-lesion progression whereas the latter more correctly incorporates variation by averaging all possibilities. Estimation of probabilities using the modified method might be biased if only a few patients remain in a trial at some timepoint. The mAug has similar properties to eAugbin with respect to precision and power when using BOR.

The extended and modified methods define progression as 20\% increase from baseline, whereas RECIST defines progression as 20\% increase from the minimum point observed. On the HORIZON II dataset, we examined the number of patients who had their best observed response being PR or CR by both of these definitions.  The number is the same for both approaches for all number of follow-up times. This indicates that considering progression as being 20\% from baseline does not substantially affect the estimation. However, we should point out that the eAugbin would be able to use the RECIST definition of progression by including a suitable indicator variable in the integrand as well as mAug by changing regions of integration of variables.

All proposed augmented binary methods involve modelling the log tumour size ratio and new-lesion progression indicators. An alternative approach is joint latent modelling of longitudinal tumour size data and the new-lesion progression. One can use a random effect model for the repeat tumour size measure, and a latent class membership for new-lesion progression. By membership, we mean a participant has probabilities of belonging to latent classes. Each class refers to the time when new-lesion progression occurs. Moreover, tumour-growth progression or new lesions appearing at a time period results in the patient’s tumour size measure being missing for all subsequent time periods. Considering this monotone missing pattern in log tumour size, the joint probability of log tumour size can be written as the product of a set of conditional probabilities of current log tumour size ratio given previous data \cite{Schafer1997}. Future work is warranted to investigate whether this more complicated methodology is worth applying.

We have only considered response end-points in this work. An increasingly commonly used phase II endpoint is progression-free-survival (PFS). Further development of the augmented binary method so that it can be applied to improve analyses of PFS is an area of current work.

\section{Software}
\label{sec5}
A package mAugbin in R is available at \href{https://sites.google.com/site/jmswason/supplementary-material}{https://sites.google.com/site/jmswason/supplementary-material} for the methods proposed in this work. The package includes extend augment binary method, which is more computationally intensive, as well as modified augmented binary method, which required an assumption for estimating the probability of response for \textit{fixed time} and for best observed response. 

\section*{Acknowledgement}
This work was supported by the Medical Research Council (grant number \texttt{MC\_UP\_1302/4}), Cancer Research UK (grant number C48553/A18113). We thank AstraZeneca for providing HORIZON II data.


\bibliography{refs}
\bibliographystyle{abbrv}

\clearpage\pagebreak\newpage

\begin{table}
\caption{\label{Table2} Mean estimated probability of response and coverage of the modified augmented binary method (mAug) in comparison with using dichotomised continuous method (Bin), augmented binary method (Augbin) and extended augmented binary method (eAugbin) using \textit{fixed time} with varying numbers of follow up times.}
\centering
\scalebox{0.90}{
\begin{tabular}{@{}ccccccccccc@{}}
\hline
Scenario  & & & \multicolumn{3}{c}{Mean of estimated probability}& \multicolumn{3}{c}{Estimated coverage}  & \multicolumn{2}{c}{Reduction in width of 95\% CI(\%)} \\
\cline{4-11}
$(\alpha, \gamma)$   & Time & True  & Bin   & Augbin/ & mAug  & Bin   & Augbin/ & mAug  & Augbin/ & mAug  \\
& & & & eAugbin & & & eAugbin& &eAugbin &\\
\hline
(-1.5, 0)   & 2    & 0.334 & 0.333 & 0.332  & 0.338 & 0.957 & 0.947  & 0.947 & 15.68  & 14.75 \\
(-2.5, 0.2) & 2    & 0.293 & 0.293 & 0.293  & 0.286 & 0.948 & 0.945  & 0.941 & 13.45  & 11.94 \\
(-1.5, 0)   & 3    & 0.318 & 0.316 & 0.314  & 0.317 & 0.953 & 0.936  & 0.949 & 12.53  & 13.26 \\
(-2.5, 0.2) & 3    & 0.450  & 0.444 & 0.443  & 0.443 & 0.954 & 0.943  & 0.948 & 14.5   & 15.28 \\
(-1.5, 0)   & 4    & 0.270  & 0.268 & 0.263  & 0.268 & 0.949 & 0.926  & 0.95  & 12.67  & 12.54 \\
(-2.5, 0.2) & 4    & 0.429 & 0.422 & 0.421 & 0.421 & 0.957 & 0.938  & 0.943 & 13.14  & 14.41\\
\hline
\end{tabular}}
\end{table}

\begin{table}
\caption{\label{Table3}Mean estimated probability of response and coverage using best observed response (BOR) without confirmation with Bin, eAugbin and mAug for maximum number of visits from 4 to 7.}
\centering
\scalebox{0.90}{
\begin{tabular}{@{}cccccccccccc@{}}
\hline
& & & & \multicolumn{3}{c}{Mean of estimated probability}& \multicolumn{3}{c}{Estimated coverage}  & \multicolumn{2}{c}{Reduction in width of 95\% CI(\%)} \\
\cline{5-12}
$(\alpha, \gamma)$ &n   & Time & True  & Bin   & eAugbin & mAug  & Bin   & eAugbin & mAug  & eAugbin & mAug  \\
\hline
(-1.5, 0)   &75          & 4    & 0.4   & 0.4   & 0.404  & 0.403 & 0.959 & 0.954  & 0.955 & 16.5   & 15.9  \\
(-1.5, 0)   &75          & 5    & 0.391 & 0.393 & 0.398  & 0.396 & 0.943 & 0.951  & 0.952 & 16.6   & 15.9  \\
(-1.5, 0)   &75          & 6    & 0.386 & 0.39  & 0.395  & 0.394 & 0.959 & 0.954  & 0.955 & 16.7   & 16    \\
(-1.5, 0)   &150         & 7    & 0.382 & 0.382 & ---    & 0.387 & 0.944 & ---    & 0.957 & ---    & 16.6\\
(-2.5, 0.2) &75       & 4    & 0.46  & 0.457 & 0.462  & 0.461 & 0.941 & 0.957  & 0.957 & 16.8   & 17.3 \\
(-2.5, 0.2) &75       & 5    & 0.452 & 0.448 & 0.454  & 0.452 & 0.954 & 0.96   & 0.96  & 18.2   & 17.2 \\
(-2.5, 0.2) &75       & 6    & 0.446 & 0.442 & 0.449  & 0.447 & 0.942 & 0.962  & 0.961 & 18.3   & 17.2 \\
(-2.5, 0.2) &150      & 7    & 0.441 & 0.441 & ---    & 0.446 & 0.95  & ---    & 0.96  & ---    & 18.3\\
\hline
\end{tabular}}
\end{table}

\begin{table}
\caption{\label{Table5} The width of 95\% CI for 3 methods using fixed time with between 2 and 5-follow up times for individual arm.}
\centering
\begin{tabular}{@{}ccccccccccccc@{}}
\hline
 & \multicolumn{4}{c}{Placebo}& \multicolumn{4}{c}{Cediranib 20 mg}  & \multicolumn{4}{c}{Cediranib 30 mg} \bigstrut \\
  \cline{2-13}
 Method $\setminus$ Time  & 2     & 3     & 4     & 5      & 2     & 3     & 4      & 5     & 2     & 3     & 4      & 5 \\
\hline
Bin      & 0.113 & 0.114 & 0.111 & 0.105  & 0.111 & 0.112 & 0.111  & 0.111 & 0.134 & 0.133 & 0.13   & 0.124 \\
eAugbin   & 0.073 & 0.074 & 0.073 & 0.07   & 0.072 & 0.075 & 0.074  & 0.064 & 0.088 & 0.088 & 0.088  & 0.085 \\
mAug     & 0.086 & 0.087 & 0.087 & 0.08   & 0.086 & 0.088 & 0.086  & 0.088 & 0.105 & 0.105 & 0.104  & 0.096\\
\hline
\end{tabular}
\end{table}

\end{document}